# Raman spectroscopy at the edges of multilayer graphene

Q. -Q. Li, X. Zhang, W. -P. Han, Y. Lu, W. Shi, J. -B. Wu, P. -H. Tan*

*State Key Laboratory of Superlattices and Microstructures, Institute of Semiconductors, Chinese Academy of Sciences, Beijing 100083, China*

**Abstract**

Edges naturally exist in a single-layer graphene sample. Similarly, individual graphene layers in a multilayer graphene sample contribute their own edges. We study the Raman spectrum at the edge of a graphene layer laid on $n$ layer graphene ($n$LG). We found that the D band observed from the edge of the top graphene layer exhibits an identical line shape to that of disordered $(n+1)$LG induced by ion-implantation. Based on the spectral features of the D and 2D bands, we identified two types of alignment configurations at the edges of mechanically-exfoliated bilayer and trilayer graphenes, whose edges are well-aligned from their optical images.

## 1. Introduction

Single layer graphene (SLG) has been intensely researched owing to its unique properties [1]. These intriguing properties also extend to multi-layer graphene (MLG). MLG samples with less than ten layers show a distinctive band structure compared with SLG. Therefore, there is an increasing interest in the physics of MLG samples, and their applications. A SLG sample will naturally feature edges, and individual graphene layers in an MLG will contribute their own edges to the overall edge of the sample. Ideally the edge of an MLG should feature good alignment of the edges of all the individual graphene layers. In reality, individual graphene layers in MLG are misaligned to a large extent with adjacent layers. The misalignment distance $h$ between the edges of two adjacent layers of MLG can be as large as a micrometer. The misalignment can also be on the nanoscale, and even zero, which cannot be distinguished by optical microscopy. The misalignment of two edges of adjacent layers in MLG is inevitable. Raman spectroscopy has been used to characterize carbon materials, and especially for studying defects [2, 3] and edges [4, 5]. To date, there have been a number of studies on the Raman spectra at the SLG and BLG edges [4, 5, 6]. However, Raman spectra at MLG edges have not yet been analyzed in detail. To understanding the fundamental physics and device applications of MLGs, it is essential to characterize the edge alignment of graphene layers in MLGs.

In this study, the Raman spectra at the edge of a graphene layer laid on $n$ layer graphene ($n$LG) are investigated. We reveal that the D band at the edge of the top graphene layer exhibits an identical line shape to that of ion-implanted ($n+1$)LG. Based on the profiles of the D and 2D bands, we identify two alignment configurations at the edges of both bilayer graphene (BLG) and trilayer graphene (TLG).

## 2. Experiments

Mechanical exfoliation of highly oriented pyrolytic graphite was used to obtain 1–4 layer graphene flakes on a Si substrate with a 90 nm layer of $SiO_2$ on the top. The number of layers in each graphene sample was identified by Raman spectroscopy [7, 8]. The edge and the misalignment distance $h$ between two adjacent layers of MLG were first characterized by optical microscopy, if $h$ was within the resolution of the microscope. The alignment configurations of MLG with $h$ smaller than the optical resolution were then identified by Raman spectroscopy. Raman spectra were measured in a backscattering geometry at room temperature using a Jobin–Yvon HR800 Raman system. The excitation wavelength was 633 nm from a He-Ne laser. The polarization of the laser beam was parallel to the edge orientation to achieve the maximum intensity for the D mode at the edge [5, 9]. A Raman line scan perpendicular to the edge orientation was used to find the maximum intensity of the D band near the edge [4]. Some graphene flakes were implanted with 90 keV carbon ($^{12}C$) ions at a fluence of $5 \times 10^{13}$ $C^+/cm^2$ by an LC-4 type system.

## 3. Results and discussion

Fig. 1a and 1b show optical images of BLG and TLG samples with well-aligned edges. However, in most cases, the edges of graphene flakes produced by mechanical exfoliation were stacked layer by layer, as shown in Fig. 1c, where SLG, BLG, TLG and four-layer graphene (4LG) can be clearly identified in the optical image. Along the dashed line in Fig.1c, the schematic diagram of the side-view atomic alignment configuration at the edges is demonstrated in Fig. 1d. To identify the complicated alignment configuration at MLG edges, we denote the well-aligned $n$LG edge as $n$LG$_{nE}$, where the subscript $E$ refers to "edge". In the general case, a specific alignment configuration at the $n$LG edge is the well-aligned edge of $m$LG (including $m = 1$) laid on $(n − m)$LG, denoted as $n$LG$_{mE}$ ($n > m$). Thus, the edge of each graphene layer in Fig. 1d can be denoted as SLG$_{1E}$, BLG$_{1E}$, TLG$_{1E}$ and 4LG$_{1E}$.

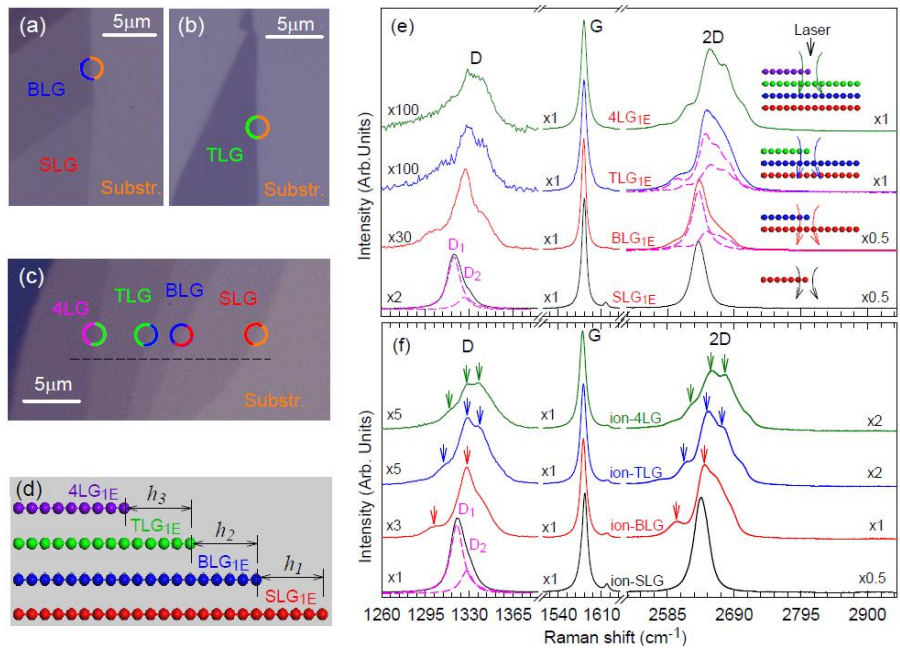

Figure 1 - Optical microscope images of BLG (a) and TLG (b) with well-aligned edges. (c) Optical microscope image of a graphene flake contained both SLG, BLG, TLG and 4LG. (d) Schematic diagram of side-view alignment configuration at edges of the graphene flake in (c). $h_i$ (i = 1, 2, 3) refer to the alignment distance between the edges of two adjacent misaligned graphene layers. (e) Raman spectra of nLG$_{1E}$ (n = 1, 2, 3, 4) as indicated by the circles in (c). The dashed lines show the 2D bands of SLG

and BLG (BLG and TLG) used to fit that at BLG$_{1E}$ (TLG$_{1E}$). (f) Raman spectra of ion-nLG (n = 1, 2, 3, 4). The arrows show the similar spectral features between the D and 2D bands of ion−nLG.

Fig. 1e depicts the Raman spectra at $n$LG$_{1E}$ edges as indicated by solid circles in Fig. 1c. The D band at SLG$_{1E}$ shows an asymmetrical profile, which can be fitted by two Lorentzian peaks (D1 and D2). This can be understood by its two different double-resonant Raman processes [10]. The D bands at BLG$_{1E}$, TLG$_{1E}$ and 4LG$_{1E}$ exhibit more complicated spectral features. The 2D bands at $n$LG$_{1E}$ ($n > 1$) are composed of bands from $n$LG and ($n − 1$)LG, as shown by dashed lines in Fig. 1e. The D mode intensity, $I$(D), at SLG$_{1E}$ is about 5 times stronger than that at BLG$_{1E}$, and $I$(D) at BLG$_{1E}$ is twice as strong as that at TLG$_{1E}$. The variation of $I$(D) with layer number is dominated by electron-phonon coupling around the K point and the optical matrix element between electronic transitions involved in the resonant Raman process of the D mode, which depends on the layer number [7].

The edge is a special case of disorder. Raman spectra of disordered $n$LG (ion-implanted $n$LG, ion-$n$LG) are depicted in Fig. 1f. The D mode profile of ion-SLG is identical to that at SLG$_{1E}$. A broad D band is observed in ion-$n$LG ($n$ = 2, 3, 4), whose profiles are very similar to the corresponding 2D band, as indicated by the arrows in Fig. 1f. $I$(2D) of ion-SLG is 3 times larger than that of ion-BLG and it is also found that the D-band profile at $n$LG$_{1E}$ resembles that of ion-$n$LG, but with much lower intensity.

$I$(D) and $I$(2D) at the $n$LG edge depends on the misalignment distance $h$. As an

example, we focus on the BLG edge with $SLG_{1E}$ and $BLG_{1E}$, when $h$ is smaller than the laser spot diameter ($R$, ~500 nm), as shown in Fig. 2a. $I(2D)$ is proportional to the area of SLG or BLG covered by the laser spot. Thus, when $h$ approaches 0 from $R$, $I(2D)$ at $BLG_{1E}$ remains constant and that at $SLG_{1E}$ decreases from a maximum to zero, as depicted in Fig. 2b. However, the D mode activated by the edge can only occur in a small area beyond which the electron (or hole) cannot be scattered by the edges [4]. There exists a critical distance $h_c$ near the edge (around 3.5 nm) [4,6]. In contrast to the 2D mode, $I(D)$ is proportional to the active length at the edge within the laser spot. As $h$ decreases from $R$ toward 0, $I(D)$ from BLG1$E$ will first remain constant until $h = h_c$, and then gradually increase up to that of $BLG_{2E}$. Meanwhile, $I(D)$ from $SLG_{1E}$ will gradually increase from 0 to a maximum until $h = h_c$, and then decrease to 0, as depicted in Fig. 2c.

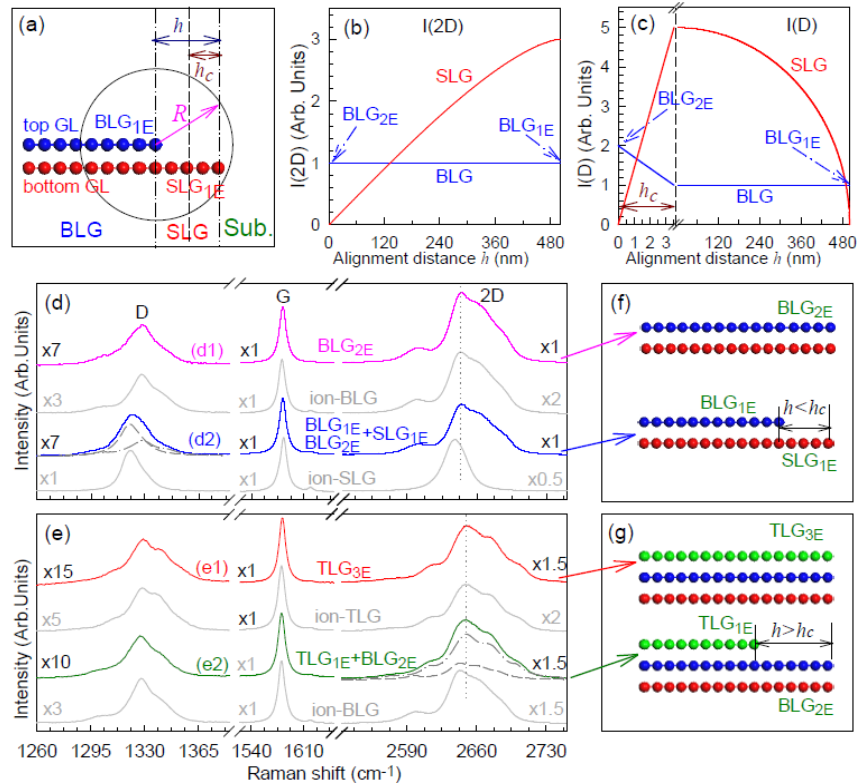

*Figure 2 - (a) Schematic illustration of a laser spot focused at the $BLG_{1E}$ of a sample, here R is radius of laser spot, h the alignment distance at the $BLG_{1E}$. I(2D) (b) and I(D) (c) of the SLG and BLG components of Raman spectrum at the BLG edge as a function of h. Typical Raman spectra ((d1) and(d2)) of two BLG samples at their likely well-aligned edges. The D bands of $SLG_{1E}$ and $BLG_{1E}$ are used to fit that at the BLG edges in (d2). Typical Raman spectra ((e1) and (e2)) of two TLG samples at their well-aligned edges. The 2D bands of ion-BLG and ion-TLG are used to fit that at TLG edges in (e2). (f) Two types of alignment configurations at the BLG edges denoted by $BLG_{2E}$ and $BLG_{1E}(BLG_{2E})+SLG_{1E}$. (g) Two types of alignment configurations at the TLG edges denoted by $TLG_{3E}$ and $TLG_{1E}+BLG_{2E}$. Raman spectra of ion-SLG, ion-BLG and ion-TLG are included in (d) and (e) for comparison. Vertical dotted lines are a guide to the eye.*

When misalignment distance *h* at *n*LG edges cannot be identified by the optical microscopy, it is possible to determine the stacking configuration at *n*LG edges by the intensity and profile of the D and 2D modes at the edges. Fig. 2d and 2e show typical Raman spectra at the edges of BLG and TLG. Both the profiles of D and 2D bands in (Fig. 2d, d1) are identical to those of ion-BLG, which suggests that the edge is well-aligned as for $BLG_{2E}$, see Fig. 2f. Similarly, the TLG edge in (Fig. 2e, e1) is also well-aligned as for $TLG_{3E}$, see Fig. 2g. For the BLG edge in (Fig. 2d, d2), its 2D band is identical to that in ion-BLG, meaning that *h* at the BLG edge is very small. The fitting in (Fig. 2d, d2) reveals that *I*(D) from SLG is twice that of BLG, indicating that *h* is even smaller than $h_c$ and its partial edge may be $BLG_{2E}$, as depicted in Fig. 2f. For the TLG edge in (Fig. 2e, e2), its 2D band can be fitted by those from BLG and TLG, suggesting a considerable area of BLG at the TLG edges. Furthermore, its D-band profile is similar to that in ion-BLG, indicating the existence of $BLG_{2E}$ at the TLG edges. Thus, the alignment configuration at this TLG edge is $TLG_{1E}+BLG_{2E}$, as depicted in Fig. 2g.

## 4. Conclusions

We presented the Raman spectra at the MLG edges. The D band at the edge of the top graphene layer laid on $n$LG shows a similar spectral profile to that of ion-implanted $(n+1)$LG. $I$(D) is proportional to the active edge length of graphene layers. The D band shows its sensitivity in identifying the alignment configuration at edges for $n$LG ($n > 1$). This study will benefit experimental research on the properties and applications of the edge states in MLGs.

**Acknowledgments**

We acknowledge support from National Nature Science Foundation of China, grants 11225421, 11474277 and 11434010.